# Straightforward measurement of thermal properties of anisotropic materials: the case of $Bi_2Se_3$ single crystal

D. Fournier[1], M. Marangolo[1], M. Eddrief[1], N.N. Kolesnikov[2], C. Fretigny[3]

[1] *Sorbonne Universités, UPMC, Univ. Paris 06, CNRS, UMR 7588, Institut des NanoSciences de Paris, 75005, Paris, France.* [2]ISSP, *Russian Acad. Sci., Chernogolovka, Moscow Distr. Russia* [3] *Laboratoire SIMM, UMR 7615, CNRS, UPMC, ESPCI, 10 rue Vauquelin, Paris, 75005, France*

**Abstract**
We show that by a simple two-steps measurement protocol one can extract the thermal properties of anisotropic insulator materials. We focus on $Bi_2Se_3$ single crystal, a layered crystal with individual sheets held together by Van der Waals force, presenting strong heat diffusion anisotropy. Since the shape of the sample does not allow experiments in the perpendicular direction the challenge is to measure the in plane and out of plane thermal properties by in plane measurements. The optical properties of $Bi_2Se_3$ being known (opaque sample for the pump and probe beams), thermoreflectance microscopy experiments can be done on the bare substrate. In this case, the lateral heat diffusion is governed by the in plane thermal diffusivity. A 100 nm gold layer is then deposited on the sample. In this second experiment it is straightforward to demonstrate that the perpendicular thermal properties of the layer rule the lateral diffusivity. The good agreement of theoretical curves obtained by thermal models with experimental data permit us to evaluate the thermal conductivity coefficients of bulk Bi2Se3.

**Introduction**
Thermal conductivity *k* is one of the leading parameters employed to describe the transport phenomena in solid state systems. Its formal definition comes from Boltzmann transport equations that clearly put into evidence the tensorial nature of *k* and incite experimentalists to measure its directional values. This is particularly important in layered system like di-chalcogenides systems ($MX_2$ with M=Mo,W; X=S, Se, Te), black phosphorus and selenides ($Bi_2Se_3$) and tellurides ($Bi_2Te_3$) of Bismuth. Indeed, in these compounds the Van der Waals bonding between layers makes the cross-plane thermal conductivity $k_z$ much lower than the in-plane one $k_r$ [see for example, Ref 1, 2]. Most of these layered systems have been extensively studied in bulk because of their thermoelectric applications since their very low values of *k* are accompanied by significant electrical conductivity (σ) and high Seebeck coefficient (S) leading to high dimensionless figure of merit $ZT = S^2 \sigma T / k$, with *T* the absolute temperature.

Inevitably, the genuine anisotropic character of *k* is lost when such materials are employed in their polycristalline form (e.g. in industrial thermoelectric applications) leading to effective *k* parameters that are affected not only by the *intrinsic* (i.e. depending of the electronic and phononic band structure) $k_z$ and $k_r$ values of the compound but also by *extrinsic* characteristics of the manufactured material (grain boundaries, disorder, defects, etc..). Interestingly, similar considerations can be done when bulk layered materials are compared with ultra thin (one or few layers) 2D materials that have attracted the attention of current research because of their electronic, spintronic and optoelectronic properties. Indeed, thermal conductivity of 2D materials is affected by *nanoscale* properties like phonon confinement, thermal diffusion into the underlying substrates and peculiar intrinsic properties like topology insulating states (ex. $Bi_2Se_3$ and $Bi_2Te_3$) and electronic band gap modifications ($MoS_2$).

As a matter of fact, thermoreflectance techniques are currently employed to extract directional *k* values [see for example Ref.3]. Basically, this kind of pump-probe experiments demands two light beams: one for the creation of the heat source and the other for the measurement of the enhanced local temperature. Light reflectance is affected by the temperature field associated to the heat diffusion, permitting to extract thermal properties by adopting convenient models. It's important to underline that $k_r$ measurements demand a quite simple set-up where thermoreflectance is measured as a function of the lateral distance between two distinct laser sources (the pump and the probe). Moreover, no absolute temperature evaluation is needed. On the other hand, $k_z$ demands either an absolute temperature evaluation in frequency mode [4-13] or picosecond acoustics and time delay measurements in time mode. [14-19] thermoreflectance based on. In the following, after a brief recall of thermoreflectance techniques, we will present a heat diffusion model in the frequency domain that permits us to envisage a new method to extract $k_z$ in bulk materials in a straightforward way without absolute temperature evaluation. Our calculations put into evidence that similarly to $k_r$, $k_z$ can be measured by lateral measurements once the bad thermal conductor sample is covered by a thin high thermal conductivity film (gold in our case). This method permitted us to measure the $k_r$ and $k_z$ values of bulk $Bi_2Se_3$, a well known thermoelectric material and topological insulator.

**Theory in the case of semi-infinite sample**

In this section, we calculate the spatio-temporal temperature field induced on the surface of a semi-infinite sample by a modulated point heat source. The sample presents a uniaxial symmetry with the symmetry axis along the normal of the surface plane. It's convenient to write the heat diffusion equation in a cylindrical coordinate system [20, 21, 22] in the following manner:

$$k_z \frac{\partial^2 T(r,z,t)}{\partial z^2} + k_r \left( \frac{\partial^2 T(r,z,t)}{\partial r^2} + \frac{1}{r} \frac{\partial T(r,z,t)}{\partial r} \right) = \rho C \frac{\partial T(r,z,t)}{\partial t} \qquad \text{Eq.1}$$

Where *T(r,z,t)* is the heat modulated temperature variation, $k_r$ and $k_z$ are the values of the thermal conductivities in- and out-of plane directions, respectively.

After a *t*-Fourier transform and an *r*-Hankel transform $T(r,z,t)$ becomes $\Theta(\lambda,z,\omega)$. In the transformed space, the solution of Eq. 1 is:

$$\Theta(\lambda, z, \omega) = A \exp\left(-\sqrt{\frac{k_r}{k_z}} \sqrt{\lambda^2 + \frac{i\omega\rho C}{k_r}} z\right) \qquad \text{Eq.2}$$

The sample region corresponds to *z* < 0. *A* is obtained from the boundary conditions given by the conservation of heat flux through the interface and by the temperature continuity at the interface (*z* = 0). If we suppose no loss in air above the sample the heat flux conservation at z = 0 reads: $-k_z \left. \frac{\partial \Theta(\lambda, z, \omega)}{\partial z} \right|_{z=0} = \Phi$, $\Phi$ being the source heat flux deposited on the surface at radial frequency ω. It is worthwhile to notice that only $k_z$ appears in the boundary condition for the anisotropic case. This gives:

$$\Theta(\lambda,z,\omega) = \frac{\Phi \exp(-\sqrt{\frac{k_r}{k_z}}\sqrt{\lambda^2 + \frac{i\omega\rho C}{k_r}}z)}{\sqrt{k_r k_z}\sqrt{\lambda^2 + \frac{i\omega\rho C}{k_r}}} \qquad \text{Eq.3}$$

The surface temperature ($z=0$) in the real space is obtained after an inverse Hankel transform:

$$T(r,0,\omega) = \int_0^\infty \frac{\Phi}{\sqrt{k_r k_z}\sqrt{\lambda^2 + \frac{i\omega\rho C}{k_r}}} J_0(\lambda r)\lambda d\lambda \qquad \text{Eq.4}$$

By integration, the analytical expression of the spherical thermal wave is obtained. It characterizes the surface heat diffusion from a point source in a semi-infinite medium:

$$T(r,0,\omega) = \frac{\Phi}{\sqrt{k_r k_z}} \exp(-\frac{r}{\mu_r})\exp(-i\frac{r}{\mu_r}) \qquad \text{Eq.5}$$

$\mu_r$ being the lateral thermal diffusion length and $D_r$ the lateral thermal diffusivity:

$$D_r = \frac{k_r}{\rho C} \quad \text{and} \quad \mu_r = \sqrt{\frac{2D_r}{\omega}}$$

Obviously, in the case of an isotropic sample $k_r = k_z = k$, Eq.5 reads:

$$T(r,0,\omega) = \frac{\Phi}{k}\exp(-\frac{r}{\mu})\exp(-i\frac{r}{\mu}) \qquad \text{Eq.5 bis}$$

A careful comparison of equations 5 (anisotropic case) and 5 bis (isotropic case) reveals that the surface temperature modulation equations just differ by the conductivity terms. It turns out that the effective conductivity in the anisotropic case is just the geometric mean of the in- and out-of-plane conductivities while the in-plane diffusivity is unchanged. In the next chapter we will show that these useful relationships are preserved in the case of an anisotropic substrate capped by an isotropic layer.

**Theory in the case of layered sample.**

Our discussion starts from the mathematical evidence that equation 1 can be written in the following way:

$$\frac{\partial^2 \tilde{T}(r,Z,t)}{\partial Z^2} + (\frac{\partial^2 \tilde{T}(r,Z,t)}{\partial r^2} + \frac{1}{r}\frac{\partial \tilde{T}(r,Z,t)}{\partial r}) = \frac{\rho C}{k_r}\frac{\partial \tilde{T}(r,Z,t)}{\partial t} \qquad \text{Eq.6}$$

Here we have supposed that the symmetry axis of the medium is parallel to the $z$ axis and we have rescaled the $z$-coordinate as $Z = \sqrt{\frac{k_z}{k_r}}z$ [23], i.e. $Z = Z_0 = \sqrt{\frac{k_{z,1}}{k_{r,1}}}z$ and $Z = Z_1 = \sqrt{\frac{k_{z,2}}{k_{r,2}}}z$ for the two different materials. Similarly to the isotropic case, it is meaningful to define a diffusion coefficient equal to $D = \dfrac{k_r}{\rho C}$. Now, we write the boundary conditions at the interface ($z = z_i$) characterized by the heat flux conservation and the equality of the temperature in the case of isotropic media:

$$k_{z,0} \dfrac{\partial T_0(r,z,t)}{\partial z}\Big|_{z=z_i} = k_{z,1} \dfrac{\partial T_1(r,z,t)}{\partial z}\Big|_{z=z_i} \text{ and } T_0(r,z_i,t) = T_1(r,z_i,t)$$

When the z-coordinate is rescaled it is possible to write *for z=$z_i$*:

$$\sqrt{k_{z,0} k_{r,0}} \dfrac{\partial \tilde{T}_0(r,Z_0,t)}{\partial Z_0} = \sqrt{k_{z,1} k_{r,1}} \dfrac{\partial \tilde{T}_1(r,Z_1,t)}{\partial Z_1} \text{ and } \tilde{T}_0(r,Z_i,t) = \tilde{T}_1(r,Z_i,t)$$

This equation written by rescaling the z-coordinate by a factor $\sqrt{\dfrac{k_z}{k_r}}$ reveals that the thermal conductivity of an anisotropic material is given by the geometrical mean, $k = \sqrt{k_r k_z}$.

In the following we will study the case of an *isotropic* layer deposited on an *anisotropic* substrate. This is the case of our experimental samples, i.e. evaporated gold on anisotropic uniaxial $Bi_2Se_3$.

First of all we recall that the surface temperature of a sample consisting of an *isotropic* layer (1) whose thickness is $h_1$ on the top of a semi-infinite *isotropic* substrate (0) heated by a modulated point source can be expressed as [5,15]:

$$T(r,0,\omega) = \int_0^\infty \dfrac{\Phi}{k_1 \sigma_1} \dfrac{k_0 \sigma_0 \text{th}(h_1 \sigma_1) + k_1 \sigma_1}{k_1 \sigma_1 \text{th}(h_1 \sigma_1) + k_0 \sigma_0} J_0(\lambda r) \lambda d\lambda \qquad \text{Eq.7}$$

where $\sigma_1 = \sqrt{\lambda^2 + \dfrac{i\omega \rho_1 C_1}{k_1}}$ is the wave vector in the layer 1 and $\sigma_0 = \sqrt{\lambda^2 + \dfrac{i\omega \rho_0 C_0}{k_0}}$ in the substrate 0.

Let us underline that in this case, no closed form solution can be found.

From the discussion above, when the substrate is anisotropic, surface temperature is obtained by replacing the substrate conductivity with its effective value in the above solution (z rescaling has no effect since the layer is supposed isotropic):

$$T(r,0,\omega) = \int_0^\infty \dfrac{\Phi}{k_1 \sigma_1} \dfrac{\sqrt{k_{0z} k_{0r}} \sqrt{\lambda^2 + \dfrac{i\omega \rho C}{k_{0r}}} \text{th}(h_1 \sigma_1) + k_1 \sigma_1}{k_1 \sigma_1 \text{th}(h_1 \sigma_1) + \sqrt{k_{0z} k_{0r}} \sqrt{\lambda^2 + \dfrac{i\omega \rho C}{k_{0r}}}} J_0(\lambda r) \lambda d\lambda \qquad \text{Eq.7bis}$$

Without getting too specific, one can say that in the case of a layered structure, the heat diffusion process is strongly depending upon the thermal properties of each component. It may be useful to describe the two asymptotic regimes reported in Fig.1

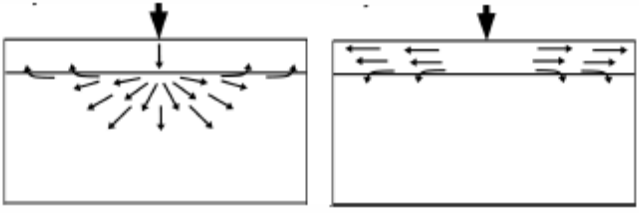

Fig.1 Lateral diffusion scheme in the case of layered structure.
Left: bad thermal conductor deposited on conductor substrate
Right: good thermal conductor on insulator.

1) If the layer has a low thermal conductivity as compared to the substrate (Fig. 1, left), the asymptotic behavior reflects the substrate lateral diffusion [8]:

$$T(r,0,\omega) \approx \frac{1}{r}\exp(-\frac{r}{\mu_0})\exp(-i\frac{r}{\mu_0})$$

2) In the opposite case (Fig. 1, right) (highly conductive layer on an insulating substrate, $k_1 \gg k_0$ and $D_1 \gg D_0$), the lateral heat diffusion corresponds to confined heat regime and the slopes of the amplitude and the phase of the asymptotic modulation temperature are neither that of the layer nor that of the substrate, as shown in references [12] and [13]:

$$T(r,0,\omega) \approx \frac{\exp(ir\sqrt{\lambda_0^2})}{\sqrt{r}} = \frac{\exp(-r(a+ib))}{\sqrt{r}}$$

with $\lambda_0^2 = -i\frac{\omega}{D_1} - \exp(\frac{i\pi}{4})\frac{k_0}{k_1}\frac{\omega}{D_0}$     Eq.8

As reported in [12,13] specific combinations of $a$ and $b$ can be written as:

$$(b^2 - a^2 + 2ab) = \frac{\omega}{D_1} = q\omega$$

Eq.9

$$h(a^2 - b^2) = \frac{k_0}{k_1}\frac{\sqrt{\omega}}{\sqrt{2}}\sqrt{\frac{1}{D_0} - \frac{1}{D_1}} \approx \frac{k_0}{k_1}\frac{\sqrt{\omega}}{\sqrt{2}}\sqrt{\frac{1}{D_0}} = p\sqrt{\omega}$$

Eq.10

The former expression, which does not depend on $h$, shows that the thermal diffusivity of the upper layer can be experimentally obtained *even if the thickness of the layer is unknown*. The latter expression attests that the thermal properties of the substrate appear as a ratio $\frac{k_0}{\sqrt{D_0}}$ which is the thermal effusivity $e_0$. If the thickness and the thermal properties of the upper layer are known, the substrate thermal effusivity $e_0 = k_0/\sqrt{D_0} = \sqrt{k_0\rho_0 C_0}$ can be measured.

Consequently, in the case of an anisotropic substrate, only the effusivity of the substrate can be deduced:

$$e_0 = \frac{k_{eff}}{\sqrt{D_{eff}}} = \frac{\sqrt{k_{0z}k_{0r}}}{\sqrt{\frac{k_{0r}}{\rho_0 C_0}}} = \sqrt{k_{0z}\rho_0 C_0} \qquad \text{Eq.11}$$

Finally, the important point of our theoretical approach is that a set of two experiments (the former on a bare anisotropic substrate and the latter on the same sample covered by an isotropic layer), permits to measure separately the sample in plane thermal diffusivity $D_r$ and the effusivity $e = \sqrt{k_z \rho C}$. Consequently, if the density and specific heat are known, one can obtain the values of $k_r$ and $k_z$.

In the following, after a brief description of the experimental set up, we will present thermal conductivities measurements of $Bi_2Se_3$ bulk samples.

**Experimental results**

**The sample**
$Bi_2Se_3$ single crystals were grown by high-pressure vertical zone melting, the method technically close to one described earlier in [24]. The initial load was pre-synthesized from elements with 99,9999 % purity in evacuated silica ampoules. The growth process was carried out in graphite crucibles with the zone movement rate of 2-4 mm/h and temperature gradient on the crystallization front of 30-40 deg. Celsius. The ambient medium was argon at 2.0 MPa pressure. The exfoliation along (0001) was carried out mechanically after cutting the crystal perpendicularly to the mentioned plane.

**The setup**
The sample surface is illuminated with an intensity modulated pump beam (532 nm laser) focused on the sample surface (around 1 μm diameter spot). The pump wavelength is chosen in order to be totally or partially absorbed by the sample. Thermal waves are then excited in the sample and monitored by the reflectivity change of another focused laser beam (probe beam). We use a 488 nm laser diode probe beam to maximize the probe sensitivity to the thermal field when the sample is covered with a gold thin film. The scanning is carried out by rotating a dichroic mirror controlled by stepper motors. A photodiode and a lock-in amplifier record the AC sample reflectivity component, in a frequency range between 1 kHz and 1MHz. Data collection and stepper motors positioning were performed by home-written software in LabVIEW ® (National Instruments). We have used very low power for the pump (532 nm laser - 1 mW) and probe beam (488 nm laser diode - 0.2 mW) to measuring bare $Bi_2Se_3$ crystal, which is opaque for the two pump and probe beams [23].

**Bulk $Bi_2Se_3$**

Figure 2 shows the experimental results (amplitude and phase) obtained by scanning the surface around the heat source point on the bare substrate. The three experimental data sets correspond to three modulation frequencies that is to say to three penetrations of the heat in the material because the thermal diffusion length is depending upon the frequency. We succeeded in obtaining a reliable set of parameters ( thermal diffusivity, spot diameter) for the fit of the three curves obtained by adopting Eq.5 ( spherical thermal waves). The good agreement obtained for each modulation frequency corroborates the thermal transport model at different probing depths ( 2-5 micrometers range). As discussed in the previous section, the measurement of the bare sample permits us to evaluate the in plane thermal diffusivity, Dr. It turns out that the Dr is estimated to 2.5 $10^{-6}$ m$^2$/s with an uncertainty of 0.25 m$^2$/s. $k_r = D_r \rho C$ can be evaluated by considering the $\rho C$ product $\rho C$ =1.4 $10^6$ J/m$^3$K [$\rho$ =7.51 g/cm$^3$ and C running [25] between 125 and 130 JK$^{-1}$mol$^{-1}$] giving k$_r$= 3.5 W/mK.

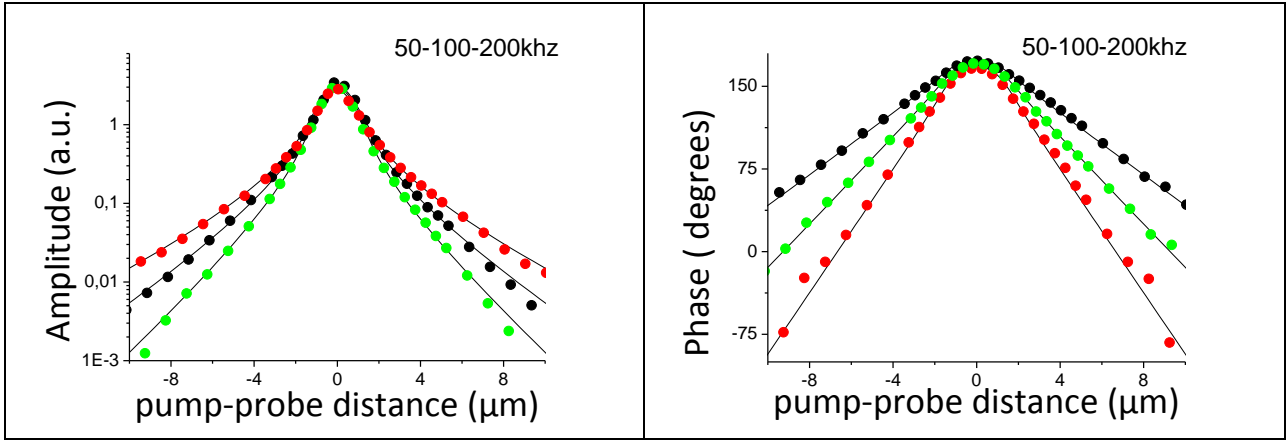

Fig.2   Thermo reflectance phase data on the Bi$_2$Se$_3$ bare sample at 50-100-200kHz
Circles: experimental data ;
Lines: theoretical model with D$_r$ = 2.5 ±0.25 $10^{-6}$ m$^2$ /s – $k_r = D_r \rho C$ = 3.5 W/mK

The sample was then covered by a 100nm evaporated Gold layer (with a hanging layer to improve adhesion). Figure 3 shows the experimental results obtained on this layered sample. As expected, the heat diffusion regime is strongly modified and differs of the previous spherical wave regime (Eq.5). Indeed, the amplitude measurements presents a $1/\sqrt{r}$ trend corresponding to a confined heat diffusion [13]. The gold layer is a very good thermal conductor deposited to a thermal insulator (the case 2 discussed in the previous chapter). By a careful evaluation of the slopes of the amplitude and phase curves [12] we obtain the effusivity of the Bi$_2$Se$_3$ crystal (Fig.3). From this value $e = \dfrac{k_{eff}}{\sqrt{D_{eff}}} = \sqrt{k_z \rho C}$ , k$_z$ is evaluated by considering the $\rho C$ product ($\rho$C=1.4$10^6$ J/m$^3$K). The out of plane thermal properties for this Bi$_2$Se$_3$ crystal are : $k_z = \dfrac{e^2}{\rho C} = 0.82 \ W/mK$ and $D_z = \dfrac{k_z}{\rho C} = 6.1 10^{-7} m^2/s$

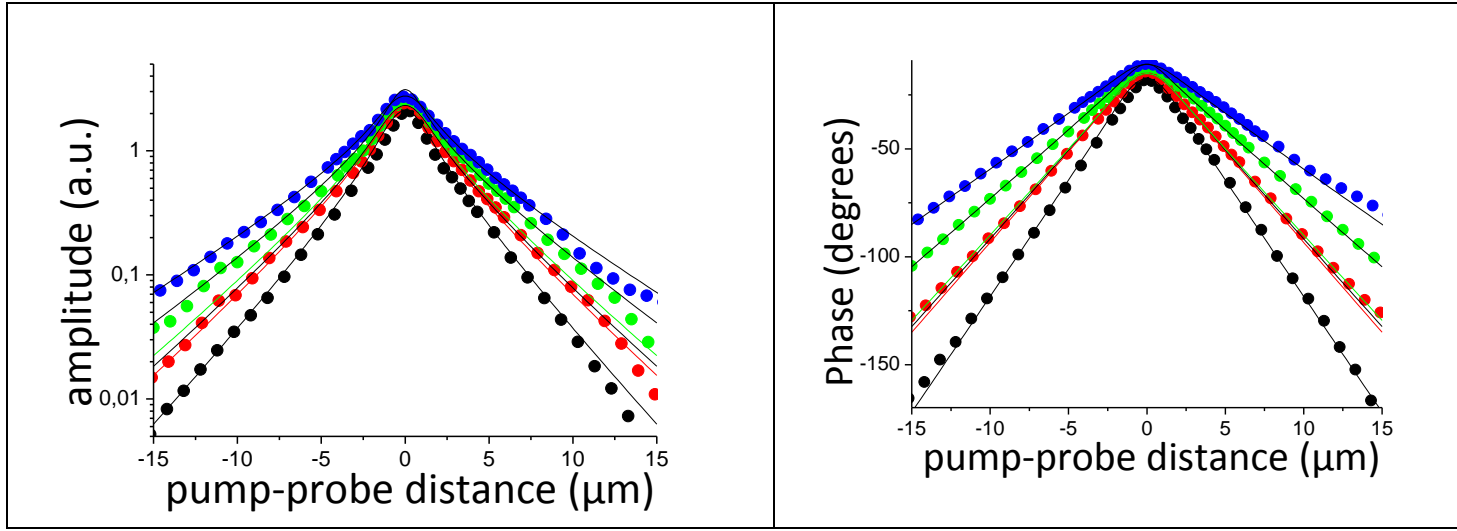

Fig 3: 100nm Gold deposited on $Bi_2Se_3$
Points: experimental data at 4 frequencies 100 kHz, 200 kHz, 400 kHz, 800 kHz
Lines: theoretical model with 100nm gold k=250 W/mK – D = 1 $10^{-4}$ $m^2$/s   $Bi_2Se_3$ effusivity =1100 ± 100 $Ws/m^2K$,

$$k_z = \frac{e^2}{\rho C} = 0.82 \pm 0.18\, W/mK - D_z = \frac{k_z}{\rho C} = 6.1 \pm 1.5\ 10^{-7}\, m^2/s$$

**Discussion**

Our experimental protocol consisting of two distinct experiments one on the bare substrate and the other one on the sample covered by a thin gold layer. The first one permitted us to evaluate the radial thermal diffusivity $D_r = 2.5\ 10^{-6}\ m^2/s$ and the second one the perpendicular thermal conductivity $k_z$= 0.9 W/mK of an anisotropic material, i.e. $Bi_2Se_3$. It turns out that a fitting procedure based on an anisotropic thermal transport model gives $D_r$= 2.5 $10^{-6}$ $m^2$/s, $k_r$= 3.5 W/mK and $k_z$= 0.82 W/mK, $D_z$ = 6.1 $10^{-7}$ $m^2$/s the density and the specific heat being close to $\rho C$ = 1.40 $10^6$ $Ws/m^3K$. These results are in good agreement with recent measurements reported in literature [26,27].

The following considerations are also worthwhile to be reported:

    1 Our measurements are not affected by the size of the pump and of the probe spots since they are based on the asymptotic part of the record.

    2 The optical absorption coefficient of the bare substrate which can lead to an eventual penetration depth of the light [23] is not a binding factor since only the asymptotic behavior of the slope of the phase is taken in consideration.

    3 An eventual interface resistance between the gold layer and $Bi_2Se_3$ due to a poor adhesion of the metallic layer plays a minor role. Indeed, an overevaluation of the interface resistance such as  R = 5 $10^{-8}$ $m^2K/W$ would lower $k_z$ to 0.65 W/mK, which is within the error bar of the measurements.

    4 We corroborated our two steps protocol by performing uncomfortable measurements on a lateral side of the layered $Bi_2Se_3$ sample, i.e. the plane containing the c- and the a-axis of the rhombohedral

structure. This permits to obtain $D_z = 8.2 \cdot 10^{-7}$ m$^2$/s and $k_z = 1.1$ W/mK, as attested by the good agreement between experimental data and fitting curves in Fig. 4.

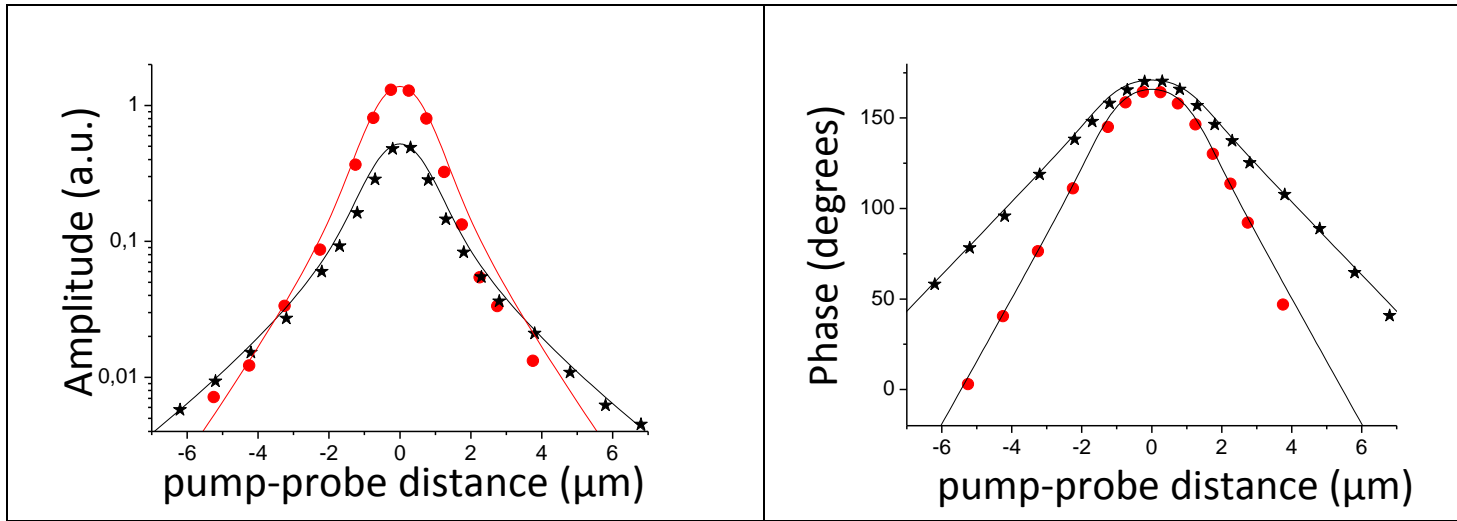

Fig 4: Experiments on two faces (top and lateral faces) of bare Bi$_2$Se$_3$ Points: experimental data at 100 kHz black points 001 axe c , red points a-axis of the rhombohedral structure
Lines: theoretical model with $D_r = 2.55 \cdot 10^{-6}$ m$^2$/s axe c and $D_r = 8.2 \cdot 10^{-7}$ m$^2$/s axe a

**Conclusion**

We proposed a very simple method to put into evidence and to measure the thermal anisotropy of a poor thermal conductor sample. This was done with experiments performed only radially. This kind of experiments is well suited when the shape or the nature of the sample do not permit to perform thermoreflectance experiments on different faces of the. We have shown that thermoreflectance measurements on the bare crystal gives the radial thermal diffusivity, while the measurements on the same sample covered by a thin gold layer gives the perpendicular thermal conductivity. We notice that sample a strong contrast of the thermal conductivity of the two materials in needed in our approach. Our experimental protocol permitted us to measure the following Bi$_2$Se$_3$ thermal properties: $D_r = 2.5 \pm 0.25 \cdot 10^{-6}$ m$^2$/s, $k_r = 3.5 \pm 0.3$ W/mK  $k_z = 0.82 \pm 0.18$ W/mK, $D_z = 6.7 \pm 10^{-7}$ m$^2$/s. These values are in good agreement with previous experimental values reported in litterature [26,27].